\documentclass[useAMS,usenatbib]{mn2e}
\usepackage{graphicx}
\usepackage{color}
\usepackage{amssymb,amsmath}
\usepackage{times}
\usepackage{psfig}

\newcommand{\g}{\gamma}

\def\spose#1{\hbox to 0pt{#1\hss}}
\newcommand\lsim{\mathrel{\spose{\lower 3pt\hbox{$\mathchar"218$}}
     \raise 2.0pt\hbox{$\mathchar"13C$}}}
\newcommand\gsim{\mathrel{\spose{\lower 3pt\hbox{$\mathchar"218$}}
     \raise 2.0pt\hbox{$\mathchar"13E$}}}

\title[TeV spectrum of PKS 1424+240 in the jet model]{Understanding the TeV emission from a distant blazar PKS 1424+240 in a Lepto-Hadronic Jet Model}
\author[Yan \& Zhang]{Dahai Yan$^{1,2}$\thanks{E--mail: yandahai555@gmail.com}, and Li Zhang$^{1}$\thanks{E--mail: lizhang@ynu.edu.cn}\\
$^1$Department of Physics, Yunnan University, Kunming, 650091 China\\
$^2$Key Laboratory of Particle Astrophysics, Institute of High Energy Physics,
Chinese Academy of Sciences, Beijing 100049, China}

\begin{document}
\pagerange{\pageref{firstpage}--\pageref{lastpage}} \pubyear{2012}

\maketitle

\label{firstpage}

\begin{abstract}
We investigate the formation of TeV spectrum of a distant blazar PKS 1424+240 residing at a redshift $z\geq0.6$ in two scenarios in the frame of a lepto-hadronic jet model taking both the uncertainties of the extragalactic background light (EBL) and its redshift into account. In the first scenario, TeV emission is attributed to the synchrotron emission of pair cascades resulting from $p\gamma$ interaction; in the second scenario, TeV emission is attributed to the proton-synchrotron emission, and an internal absorption due to interaction with the photons around the jet is included. The results show that in the first scenario the 68\% upper limit of its redshift within which this scenario can explain the VERITAS TeV spectrum in 2009 well is $\sim0.75$; in the second scenario, this upper limit of the redshift becomes $\sim1.03$. However, the second scenario can be excluded because it requires an unreasonable photon field around the jet with a luminosity of $\sim10^{43}\rm \ erg\ s^{-1}$. In conclusion, the jet model can explain its TeV spectrum with a low EBL density if $0.6<z<0.75$.
\end{abstract}


\section{Introduction}

PKS 1424+240 is classified as a BL Lac Object (BL Lacs) because of its weak emission-line feature.
Very high energy (VHE) gamma-rays from PKS 1424+240 have been detected by VERITAS \citep{Acciari2010,Archambault14} and MAGIC \citep{Aleks2014}. The intrinsic TeV emission from a blazar is expected to be attenuated through interaction with the ultraviolet-infrared photons of extragalactic background light (EBL). This attenuation depends on the distance from the source to the Earth and the EBL density. However, as for many BL Lacs, the redshift of PKS 1424+240 is still uncertain. \citet{Furniss13} presented a firm redshift lower limit of $z\geq0.6035$ for PKS 1424+240 according to the observations of Ly$\beta$ and Ly$\gamma$ absorption. Based on its gamma-ray spectra, the redshift upper limits of $z\leq0.81$, $z\leq1.00$ and $z\leq1.19$ were reported by \citet{Aleks2014}, \citet{Scully}, and \citet{yang}, respectively. A photometric redshift upper limit of $z\leq1.10$ was reported by \citet{rau}. On the other hand, the EBL density is also not measured well so far \citep[e.g.,][]{Hauser01}. Many models have been proposed to estimate the EBL intensity \citep[e.g.,][]{stecker06,Franceschini,Finke10,Kneiske10,Dominguez11,Gilmore12,Inoue13}.
The current gamma-ray observations can put strong constraints on the EBL intensity in optical-UV frequency range \citep[e.g.,][]{Ackermann2012,Yuan12,Abramowski2013}, and the relevant results indicate that the EBL intensity is closer to the lower limit (e.g., the EBL model of \citet{Franceschini}) of the EBL models.

One important and interesting issue is to explain the gamma-ray emission from the distant blazar PKS 1424+240, especially its hard TeV spectrum. Generally speaking, two kinds of models have been proposed to study the origins of the gamma-ray emissions of blazars: jet and non-jet models. The jet models can be divided into leptonic and hadronic models. In the leptonic jet models, the gamma-rays are produced through inverse Compton scattering of low-energy photons by the high energy electrons in the jet, including synchrotron-self Compton (SSC)  \citep[e.g.,][]{Maraschi1992,Yan14} and external Compton models \citep[e.g.,][]{Dermer93,Sikora1994,Yan12}. In the hadronic jet models, the high energy photons come from the proton-synchrotron emission or $p\gamma$ interaction in the jet \citep{Aharonian2000,mucke2003,bottcher09,bottcher13,Dermer12,dimit2012,dimit2014,masti13,Weidinger2014}.
In the non-jet models, the gamma-rays are produced in the propagation of ultra-high-energy-cosmic-rays (UHECRs) in the intergalactic space \citep[e.g.,][]{essey10a,essey10b,essey14,Takami13,Aharonian2013}. Moreover, some novel physical phenomena such as oscillations of photons into axion-like particles are proposed to mitigate the serious EBL-absorption of TeV photons emitted by a distant blazar \citep[e.g.,][]{Angelis,Horns,Harris,Rubtsov,Tavecchio14}.
For PKS 1424+240, \cite{Acciari2010} modeled its multi-wavelength spectral energy distribution (SED) with SSC model; however, the highest gamma-ray data cannot be explained in this model due to the strong EBL absorption. \cite{essey14} showed that the VERITAS TeV spectrum in 2009 reported in \cite{Acciari2010} could be explained well in the non-jet model with the redshift up to 1.3. \citet{Aleks2014} presented that the TeV spectrum of PKS 1424+240 derived from the observations of MAGIC in 2010 and 2011 could be explained well by a two-component SSC model when its redshift $z\sim0.6$.
We shall point out that the primary gamma-ray emission from the jet would play an important role in the non-jet models (see the results in \cite{essey14}).

\citet{Archambault14} reported the TeV spectrum of PKS 1424+240 derived from the deep VERITAS observations in 2009 and 2013. The EBL-corrected TeV spectra with $z=0.60$ and a low density EBL model show a possible indication of spectral hardening at the highest energies, which challenges the leptonic models .
To better understand the origin of the TeV emission of PKS 1424+240, the first step is to study the TeV photons produced in its jet in detail.

In this work, we investigate the formation of TeV spectrum of PKS 1424+240 in a lepto-hadronic hybrid jet model. In this model, there are two scenarios according to the production of the TeV photons, which are described in Section~\ref{secmodel}. The effects of the EBL and the redshift uncertainties are considered in this model. Our results are presented in Section~\ref{result}, and our conclusions are given in Section~\ref{cd}.

\section{The model}
\label{secmodel}

\begin{figure}
  \begin{center}
  \begin{tabular}{c}
\hspace{-0.90cm}
     \includegraphics[width=90mm,height=80mm]{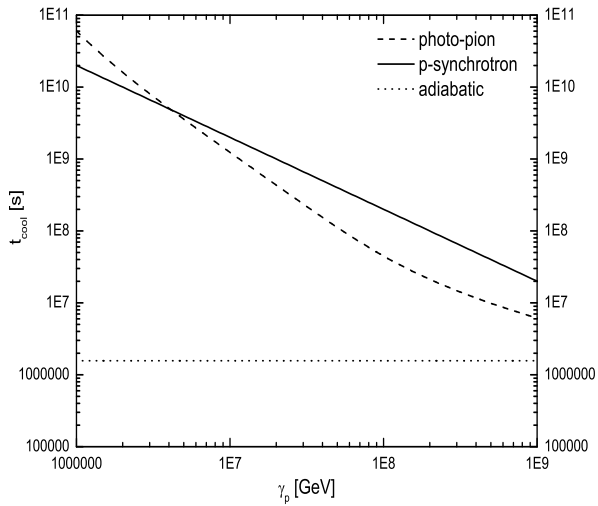}
\end{tabular}
  \end{center}
\caption{Proton cooling timescales in the blob rest frame. The parameters for model A in Table~\ref{modelp} are used.} \label{cooltime}
\end{figure}

\begin{figure}
  \begin{center}
  \begin{tabular}{c}
\hspace{-0.90cm}
     \includegraphics[width=90mm,height=80mm]{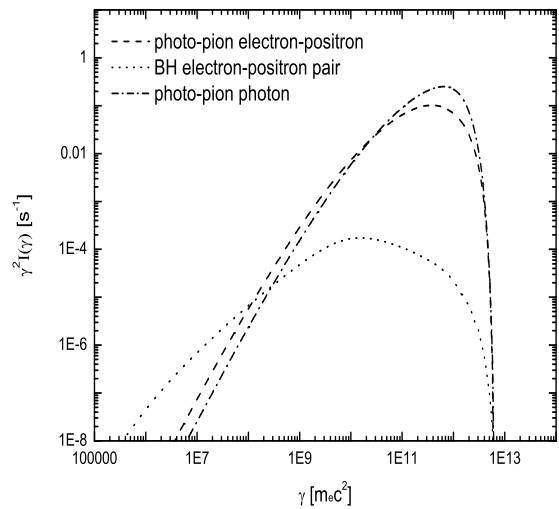}
\end{tabular}
  \end{center}
\caption{Electron-positron and photon production rates in $p\gamma$ interaction in the blob frame. The parameters for model A in Table~\ref{modelp} are used.} \label{electrondis}
\end{figure}

\begin{figure}
  \begin{center}
  \begin{tabular}{c}
\hspace{-0.90cm}
     \includegraphics[width=90mm,height=80mm]{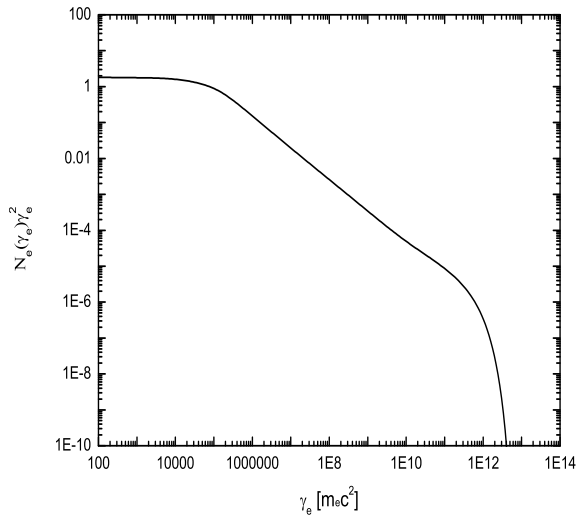}
\end{tabular}
  \end{center}
\caption{Pair cascades distribution. The parameters for model A in Table~\ref{modelp} are used.} \label{casEED}
\end{figure}

Since high energy electrons and protons may coexist in the jet of a blazar, both leptonic and hadronic emission processes should be considered \citep[e.g.,][]{mucke2003,masti13,bottcher13} in such a case, which is called as lepto-hadronic blazar jet model.
In the lepto-hadronic model, it is assumed that both high energy electrons and protons are injected into a homogeneous and spherical emission region (blob) with radius $R_{\rm b}$, magnetic-field strength $B$, and Doppler factor $\delta_{\rm D}$. In the emission region, synchrotron and SSC emissions are the two main emission processes for the primary high energy electrons; meanwhile, proton-synchrotron emission and $p\gamma$ interaction are responsible for the production of high energy photons in the hadronic processes. The synchrotron emission photons of primary high energy electrons not only provide the target photons for $p\gamma$ interaction, but also provide the low energy photons for $\gamma\gamma$ interaction. Therefore, the ultrahigh energy (UHE) photons from the $\pi^0$ decay and synchrotron emission of the first generation $e^{\pm}$ pairs produced in the $p\gamma$ interaction interact with the low energy target photons, and then the pair cascades are induced; consequently, the final gamma-rays escaping from the blob are dominated by the synchrotron emission photons of the pair cascades and protons. Briefly, this model includes the following processes: (1) synchrotron and SSC emissions of primary electrons; (2) proton-photon pion production; (3) proton synchrotron emission; (4) proton-photon pair (Bethe-Heitler or BH pair) production; (5) photon-photon pair production; and (6) synchrotron emission of UHE-photons-induced pair cascades.

\section{Results}
\label{result}

\begin{figure*}
  \begin{center}
  \begin{tabular}{cc}
\hspace{-0.90cm}
     \includegraphics[width=85mm,height=85mm]{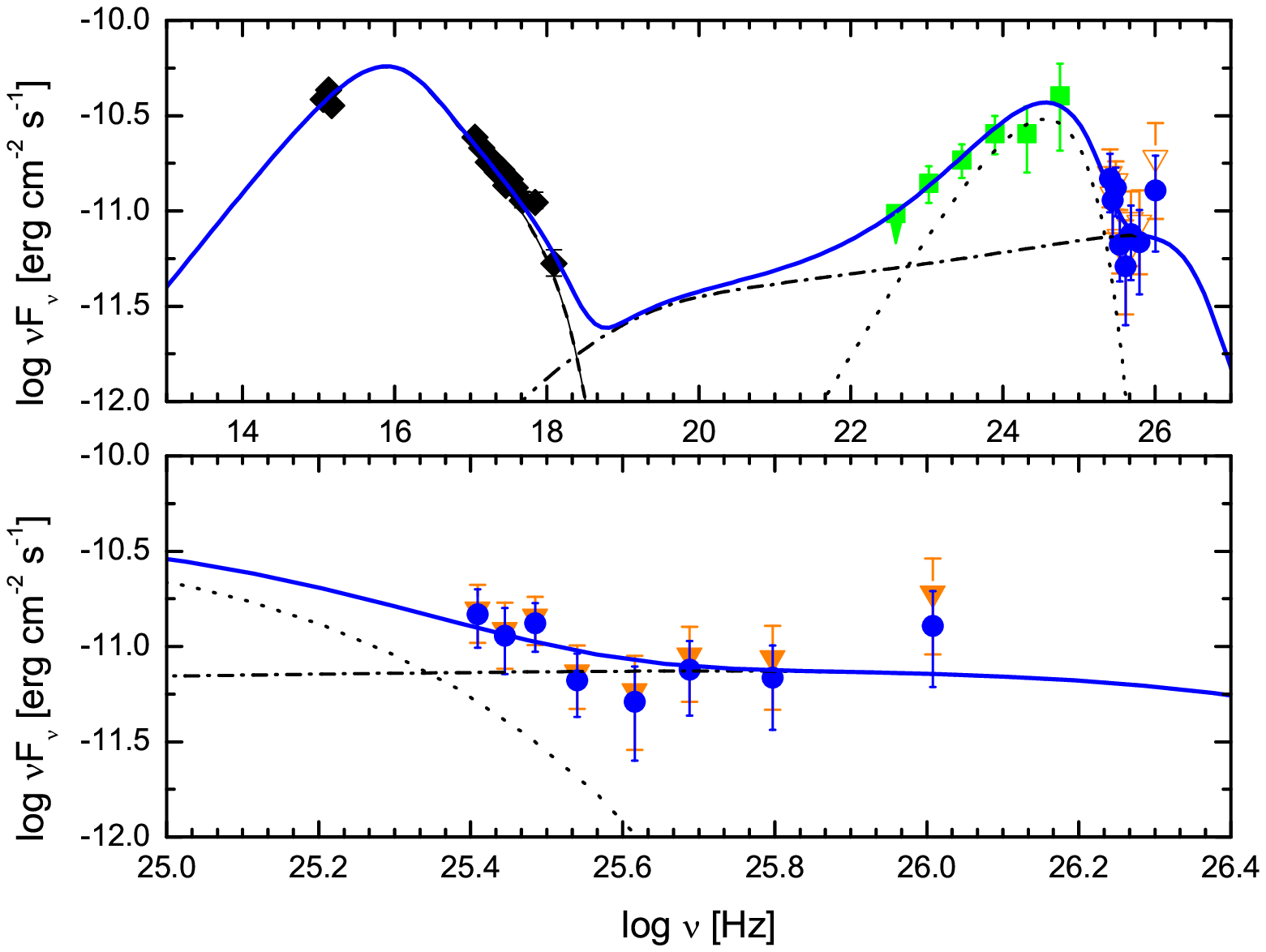} &
\hspace{-0.90cm}
     \includegraphics[width=85mm,height=85mm]{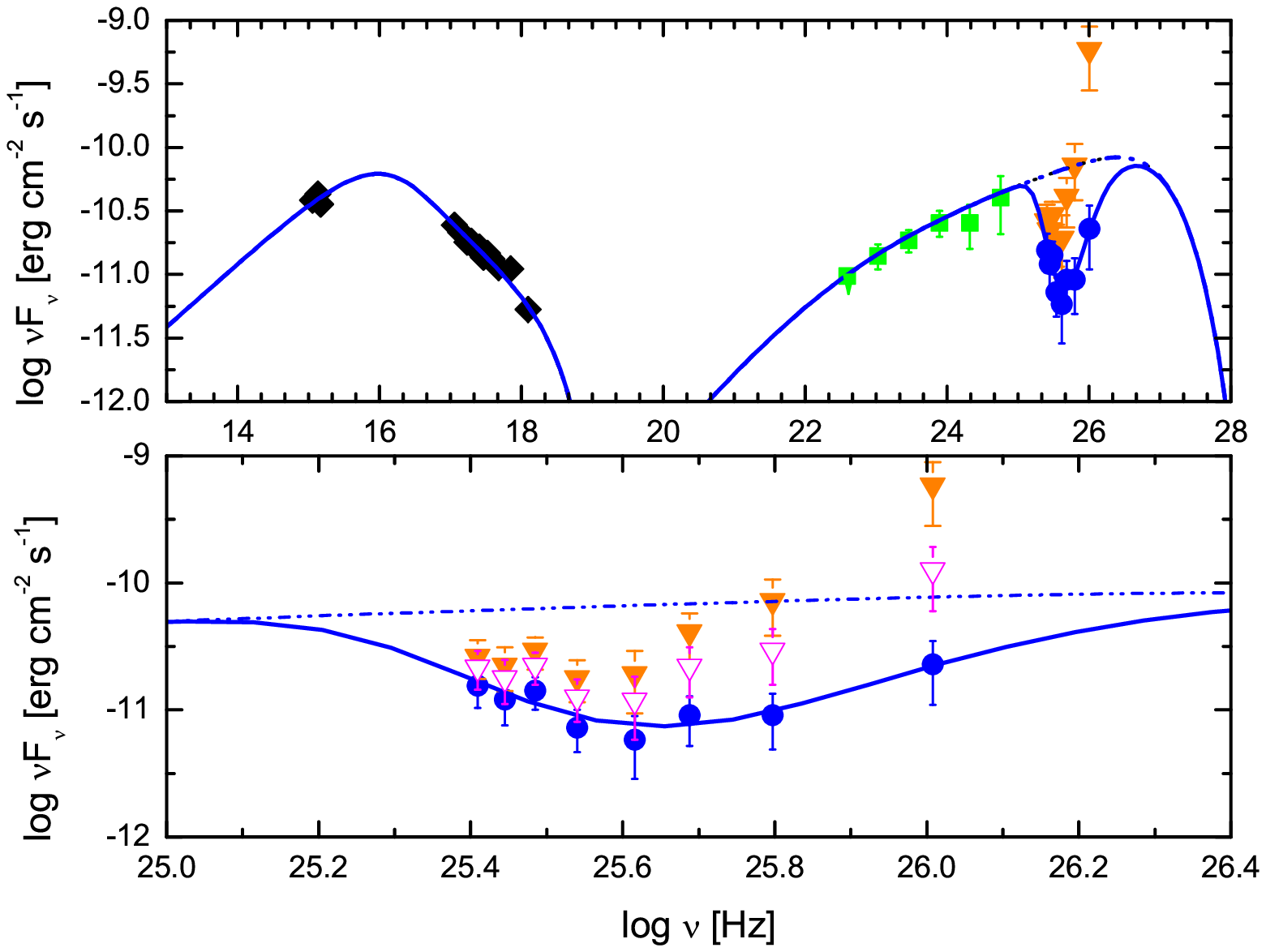}
\end{tabular}
  \end{center}
\caption{Modeling SED results in model A (left) and model B (right). Bottom panels are the TeV windows of top panels. Left panel, dashed line: synchrotron emission of primary electrons; dotted line: synchrotron emission of protons; dot-dashed line: synchrotron emission of $e^{\pm}$ pairs cascade; solid line: the sum of all kinds of emissions.
TeV data are EBL-corrected; filled circles: with $z=0.69$ and $\tau_{\rm f}=0.61$; triangles: with $z=0.75$ and $\tau_{\rm f}=0.61$ . Right panel, double-dot-dashed line: synchrotron emission of proton; solid line: synchrotron emission of proton with an internal absorption. TeV data are EBL-corrected; filled circles: with $z=0.65$ and $\tau_{\rm f}=0.78$; filled triangles:  with $z=1.03$ and $\tau_{\rm f}=0.78$; open triangles: with $z=1.03$ but $\tau_{\rm f}=0.61$. } \label{SED}
\end{figure*}

\begin{table*}
\begin{center}
\caption{Model parameters for the results of Fig.~\ref{SED} in the blob rest frame. }\label{modelp}
\begin{tabular}{lcccccccccccccc}
 \hline
Model & $B$ & $\delta_{\rm D}$ &
$R_{\rm b}$ & $\eta_{\rm esc}$ & $\gamma_{\rm e,1}$ & $\gamma_{\rm e,2}$ & $s_{\rm e}$ & $\gamma_{\rm p,1}$ & $\gamma_{\rm p,2}$ & $s_{\rm p}$ & $U_p/U_B$ & $U_e/U_B$\\
 & (G) & & ($10^{16}$\ cm) & & (GeV) & (GeV) & & (GeV) & (GeV) & & \\
 \hline
Model A & 15 & 30 & 0.47 & 10 & 1.4 & 17.9 & 3.0 & 51.1 & $3.5\times10^{9}$ & 1.6 & 1.4 & $7.7\times10^{-4}$ \\
Model B & 50 & 30 & 4.4  & 10 & 0.8 & 12.8 & 3.1 & 51.1 & $3.0\times10^{10}$ & 1.6 & $7.7\times10^{-5}$ & $1.3\times10^{-6}$\\
\hline
\end{tabular}
\end{center}
\end{table*}

Several kinds of particles energy distributions (PED) have been used to fit the SEDs of blazars \citep[e.g.,][]{Yan13,Zhou14}. Here we do not consider the acceleration processes, and just assume that particles are injected with a power-law distribution, i. e.,
\begin{equation}
Q_{\rm e/p}(\gamma_{\rm e/p})=k_{\rm e/p}\gamma^{-s_{\rm e/p}}, \ \ \gamma_{\rm e/p, 1}\leq\gamma_{\rm e/p}\leq\gamma_{\rm e/p, 2}\;,
\label{Eq1}
\end{equation}
where $k_{\rm e/p}$ are the electron/proton injection constants; $\gamma_{\rm e/p, 1}$ and $\gamma_{\rm e/p, 2}$ are the minimum and maximum values of electron/proton Lorentz factor. A steady-state solution to the particle continuity equation is used to calculate the emission spectra \citep[e.g.,][]{Dermerbook,bottcher13}, where continuous injection is balanced by the cooling and escape of the particles.

For the injected electrons, we assume an energy-independent escape timescale
$t_{\rm esc}=\eta_{\rm esc} R_{\rm b}/c$, where $\eta_{\rm esc}>1$ is a constant.
In the lepto-hadronic model, the strength of magnetic field is so large that the radiative cooling of the electrons is dominated by the synchrotron emission. When the escape timescale equals to the cooling timescale, a critical energy $\gamma_{\rm c}$ is determined. Depending on whether $\gamma_{\rm c}$ is larger than $\gamma_{\rm e, 1}$ or not, there are two regimes: slow-cooling and fast-cooling regimes. In the
slow-cooling regime with $\g_{e,1} < \g_c$ , the steady-state electron distribution can be approximated as \citep[e.g.,][]{Dermerbook,Finke13,bottcher13}
\begin{equation}
N_e(\g_e) \approx k_{\rm e} \left\{ \begin{array}{ll}
  (\g_e/\g_c)^{-s_e}   & \g_{e,1} \leq \g_e \leq \g_c \\
  (\g_e/\g_c)^{-s_e-1} & \g_c < \g_e < \g_{e,2}
      \end{array}
\right. \ .
\end{equation}
But in the fast-cooling regime with $\g_{e,1} > \g_c$,  the steady-state electron distribution is given by
\begin{equation}
\label{fast_cool_elec}
N_e(\g_e) \approx k_{\rm e}  \left\{ \begin{array}{ll}
  (\g_e/\g_{e,1})^{-2}   & \g_c \leq \g_e \leq \g_{e,1} \\
  (\g_e/\g_{e,1})^{-s_e-1} & \g_{e,1} < \g_e < \g_{e,2}
      \end{array}
\right. \ .
\end{equation}
The above electron energy distributions are used to calculate the synchrotron-SSC spectra of the primary electrons with the methods given in \citet{Finke08}.

For a given injection rate, the steady-state proton energy distribution is given by \citep[e.g.,][]{bottcher13}
 \begin{equation}
 N_p(\gamma_p)=\frac{1}{|\dot{\gamma}_p|}\int_{\gamma_p}^\infty Q_p(\gamma^{\prime}_p)d\gamma^{\prime}_p \ ,
  \end{equation}
where the escape term is negligible in the calculation of high energy emissions from the protons in the blob, because it only affects the lowest-energy protons which do not substantially contribute to the radiation \citep{bottcher13}. $|\dot{\gamma}_p|$ is the energy losses of the protons and includes radiative energy losses (proton-synchrotron and $p\gamma$-interaction cooling) and adiabatic energy loss of the protons. The proton-synchrotron cooling rate is
 \begin{equation}
|\dot{\gamma}_{p,~\rm syn}|=2.3\times 10^{5}\left(\frac{B}{100\ {\rm G}}\right)^2\left(\frac{E_p}{10^{19}\ {\rm eV}}\right)^2\ {\rm s^{-1}} \;.
  \end{equation}
The photo-pion energy loss is given as \citep{Aharonian2000}
 \begin{equation}
 |\dot{\gamma}_{p,~\rm p\gamma}|=3\times 10^{20}\frac{\langle\sigma_{p\gamma}f\rangle}{\rm cm^2} \frac{n_{\rm syn}(\epsilon_0)\epsilon_0}{\rm cm^{-3}} \frac{E_p}{10^{19}\ {\rm eV}}\ {\rm s^{-1}} \;,
  \end{equation}
where $\langle\sigma_{p\gamma}f\rangle\approx10^{-28}\ {\rm cm^2}$ is the inelasticity-weighted $p\gamma$ interaction cross section and $n_{\rm syn}(\epsilon)$ is the synchrotron photons spectral density emitted by primary electrons and $\epsilon_0=5.9\times10^{-8}(\frac{E_p}{10^{19}\ {\rm eV}})^{-1}$. The adiabatic cooling rate is
   \begin{equation}
 |\dot{\gamma}_{p,~\rm ab}|= 9.0\times10^{5}\frac{E_p}{10^{19}\ {\rm eV}}\left(\frac{R_b}{10^{15}\ {\rm cm}}\right)^{-1}\cdot\frac{1}{\delta_{\rm D}}\ {\rm s^{-1}} \;,
  \end{equation}
 where we assume the opening angle of the conical jet $\theta_j$ to be $1/\delta_{\rm D}$.
 Finally, $|\dot{\gamma}_p|$ can be written as
 \begin{equation}
 |\dot{\gamma}_p|=|\dot{\gamma}_{p,~\rm syn}|+|\dot{\gamma}_{p,~\rm p\gamma}|+|\dot{\gamma}_{p,~\rm ab}| \;.
  \end{equation}
It can be seen in Fig.~\ref{cooltime} that the adiabatic energy loss of proton significantly dominates over the radiative losses. This fact means that the steady-state proton distribution is determined by $|\dot{\gamma}_{p,~\rm ab}|$ for a given proton injected rate.

With the above proton energy distribution, the proton-synchrotron spectrum is calculated by the method used to calculate the electron synchrotron radiation, but the Larmor frequency is re-scaled by a factor of 1830 \citep[e.g.,][]{Aharonian2000}, and we restrict the proton's Larmor radius to be smaller than the radius of emission region. The final products resulted from $p\gamma$ interaction (UHE photons, the first generation $e^{\pm}$, and neutrinos) are calculated by using the methods given by \citet{Kelner}, and the target photons are the synchrotron photons of primary electrons. The energy distribution of UHE-photon-induced pair cascades and their synchrotron emission are evaluated by using the semi-analytical method given by \citet{bottcher13}. The $\gamma\gamma$ absorption optical depth for high energy photons having dimensionless energy $\epsilon_{1}$ in the blob rest frame, $\tau_{\gamma\gamma,~\rm syn}$ , due to interaction with the internal synchrotron radiation field, is calculated as \citep[e.g.,][]{Finke08}
\begin{equation}
\tau_{\gamma\gamma,~\rm syn} (\epsilon_{1} ) = \frac{3R_{\rm b}\sigma_{\rm T}}{8\epsilon^2_{1}}\int^{\infty}_{1/\epsilon_{1}}d\epsilon n_{\rm syn}(\epsilon)\bar{\phi}(\epsilon_1\epsilon)\epsilon^{-2}\ ,
\end{equation}
where the function $\bar{\phi}(\epsilon_1\epsilon)$ can be found in \citet{Gould67}, and $\sigma_{\rm T}$ is Thomson cross section. This absorption will modify the high-energy spectrum by the factor
$$\frac{1-e^{-\tau_{\gamma\gamma,~\rm syn}(\epsilon_{1} )}}{\tau_{\gamma\gamma,~\rm syn}(\epsilon_{1} )}\ .$$

In Fig.~\ref{electrondis} the electron-position and photon production rates in $p\gamma$ interaction are shown. One can see that at $\gamma_{\rm p}\gtrsim50$\ TeV energies the electron production rate of photo-pion process dominates over that of BH pair process. At lower energies the latter is dominant; however, this fact does not mean the proton cooling due to BH pairs production becomes dominant, because at low energies the proton's adiabatic energy loss is significantly dominant (see Fig.~\ref{cooltime}). Based on the results in Figs.~\ref{cooltime} and \ref{electrondis}, one can find that BH pair production is negligible for very high energy protons (e.g., $E_p\sim10^{18}\ $eV), which has been pointed out by the previous studies \citep[e.g.,][]{Aharonian2000,mucke2003}. For relative low-energy protons (e.g., $E_p\sim10^{15,16}\ $eV), BH pair production is an important source of electron-positrons, which could produce significant contribution to the SED in the hard X-ray/soft gamma-ray energy range \citep{Petropoulou}.
The energy of the first generation electron-positrons and photons produced in $p\gamma$ interaction are transferred to pair cascades and then more low energy electrons are produced (Fig.~\ref{casEED}). Below $\sim0.1$\ TeV, the distribution of the spectral index of pair cascades is $\sim-2$, and becomes $\sim-3$ at higher energies. In our calculations, it is found that the first generation electron-positrons produced in $p\gamma$ interaction have slight effect on the distribution of pair cascades, which denotes that the electrons resulted from $\gamma\gamma$ interaction are the main source of  pair cascades.

The synchrotron emissions from intermediate decay products (muons and pions) are neglected in this model, also see the recent papers \citet{bottcher13} and \citet{Weidinger2014}, implying that the decay timescales of the intermediate products is shorter than their synchrotron timescales. Although several authors argued that this effect of not considering the radiation of the intermediate particles is small \citep{mucke2003,Weidinger2014},
it should be cautious to apply this model to the situation with both high magnetic field and high electron as well as proton densities.

With different model parameters, the TeV emission produced in this lepto-hadronic jet model could be attributed to the synchrotron emission of UHE-photon-induced pair cascades (labeled as model A) or to the synchrotron emission of the protons (labeled as model B). The most basic differences between models A and B are the injected densities of electrons and protons.
The injected densities of electrons and protons in model A are much higher than those in model B (see Table 1), resulting in higher $p\gamma$ interaction rate in model A.
In model B, an internal absorption of gamma-rays due to interaction with the soft photon field around the jet is introduced just as done in \citet{Aharonian2008} and \cite{Zach11}. This internal $\gamma\gamma$ optical depth for the high energy photon having dimensionless energy $\epsilon_{1}$ in the host galaxy frame is estimated as \citep{Zach11}
\begin{equation}
\tau_{\rm int}(\epsilon_{1})\simeq0.2\sigma_{\rm T}R n_{\rm ph}(3.5/\epsilon_{1}) \ ,
\end{equation}
where $R$ is the travel distance of the gamma-rays in the photon field (broad line region or dust torus), and the soft photon field around the jet $n_{\rm ph}(\epsilon)$ is assumed to be a diluted black body spectrum that is characterized by a temperature $T$ and a diluted energy density $U_0$. This absorption will modify the high-energy spectrum by the factor $e^{-\tau_{\rm int}(\epsilon_{1})}$.

We use the cosmology parameters $H_{0}=70\rm \ km\ s^{-1}\ Mpc^{-1}$, $\Omega_{\rm m}=0.3$ and $\Omega_{\rm \Lambda}=0.7$ in our calculations.

\begin{table}
\begin{center}
\caption{The constraints on $z$ and $\tau_{\rm f}$ derived from the fits to the observed TeV spectrum. }\label{TeV}
\begin{tabular}{lcccccccccccccc}
 \hline
Model & $z$ & $\tau_{\rm f}$ & $\chi^2$/dof\\
 \hline
     model A &     0.69  & 0.61  & 0.61\\
  68\% limit &  $<0.75$  & $<0.83$   \\
  \hline
     model B &     0.65  & 0.78  & 0.38\\
  68\% limit &     $<1.03$ &  $<1.00$\\
\hline
\end{tabular}
\end{center}
\end{table}

\begin{figure*}
  \begin{center}
  \begin{tabular}{cc}
\hspace{-0.90cm}
     \includegraphics[width=80mm,height=60mm]{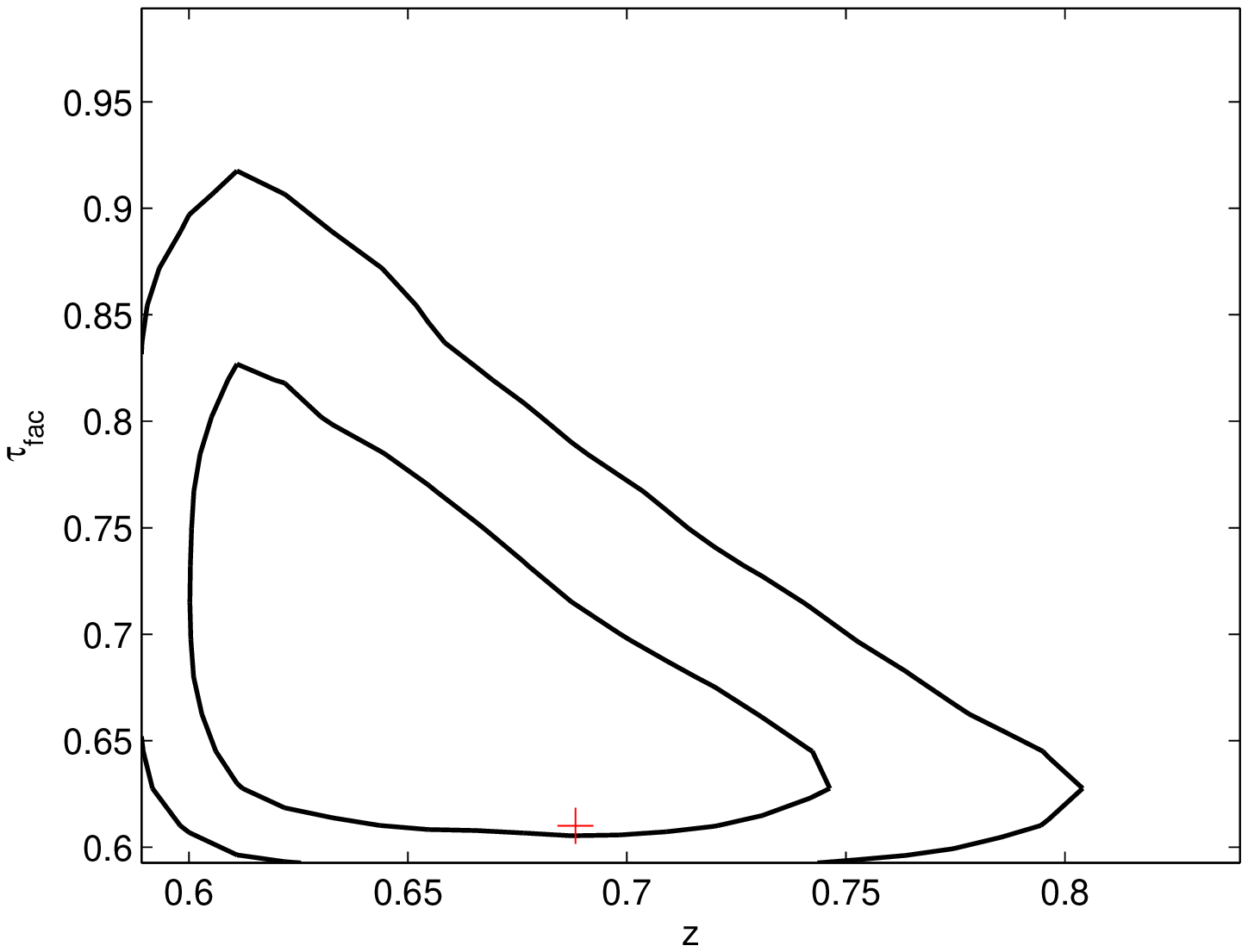} &
\hspace{-0.90cm}
     \includegraphics[width=80mm,height=60mm]{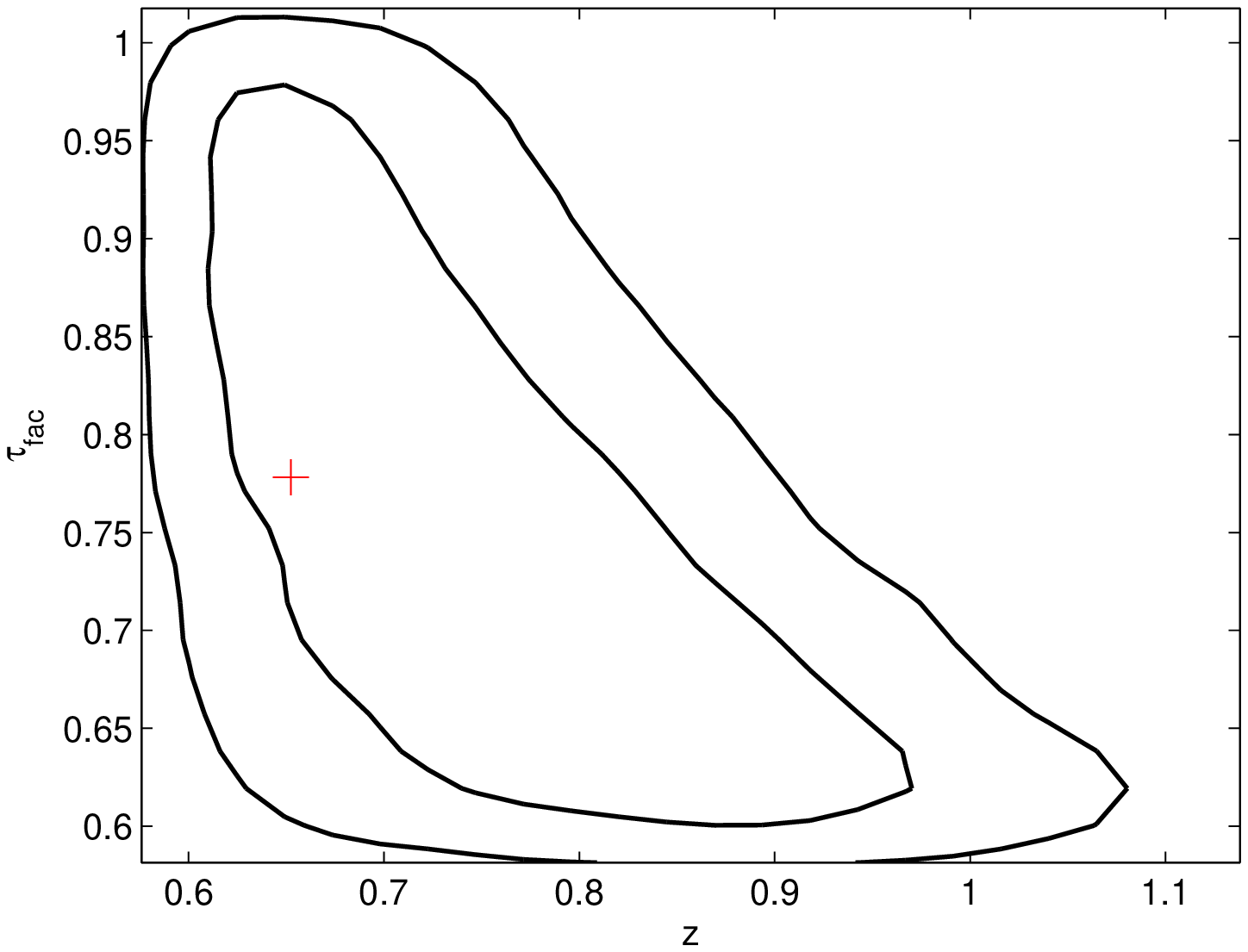}
\end{tabular}
  \end{center}
\caption{68\% and 95\% limits of $z$ and $\tau_{\rm f}$. Left: model A; right: model B with $T=1.09^{+0.55}_{-0.19}\ $eV and $R=2.09^{+0.17}_{-0.69}\ $pc. Crosses are the best fit values.} \label{limit}
\end{figure*}

\begin{figure}
  \begin{center}
  \begin{tabular}{c}
\hspace{-0.90cm}
     \includegraphics[width=90mm,height=80mm]{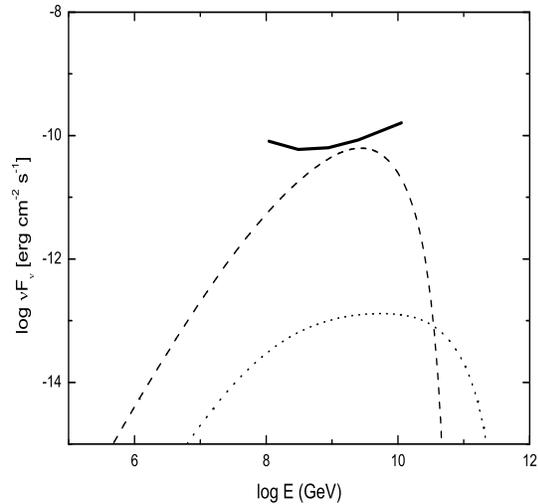}
\end{tabular}
  \end{center}
\caption{Predicted spectra of the neutrinos (electron neutrinos + electron antineutrinos + muon neutrinos + muon antineutrinos ) in models A (dashed line) and B (dotted line). The solid line is the three years' sensitivity of Askaryan Radio Array, which is taken from \citet{murase14}. } \label{neutrino}
\end{figure}

\citet{Archambault14} reported the TeV spectra of PKS 1424+240 derived from the deep VARITAS observations in 2009 and 2013 as well as the contemporaneous {\it Fermi }-Large-Area-Telescope (LAT)-detected GeV spectra and {\it Swift}-detected optical-UV spectra, where multiwavelength spectrum in 2009 is slightly above that in 2013. They have shown that no strong variability was found from their observations. \citet{Aleks2014} reported that the corrected MAGIC spectra in  2010 and 2011 with $z=0.6$ and the EBL model given by \citet{Franceschini}  are flat without any apparent turn-down up to 400 GeV. It can be found that the MAGIC spectrum in 2009 is consistent with the VERITAS spectrum in 2009. Here the VERITAS spectrum in 2009 \citep{Archambault14} is adopted because of its more data points.

First, assuming $z=0.6$, we model the SEDs of PKS 1424+240 covering from optical to GeV gamma-rays energy range to obtain the intrinsic TeV spectra in models A and B. The model results are shown in Fig.~\ref{SED} and the model parameters are listed in Table~\ref{modelp}.
It can be found that both models A and B can reproduce the observed SED from optical to GeV band well. The SSC emission from the primary electrons is negligible in the two models. In model A the synchrotron emission of protons is dominant at GeV band and the synchrotron emission of pair cascades is dominant at TeV band. In model B the synchrotron emission of protons is dominant at GeV-TeV band. Obviously, the intrinsic TeV spectra predicted by the two models are different (see Fig.~\ref{SED}). Because of the contribution from pair cascades, a spectral harding at sub-TeV band appears in model A, and then the TeV spectrum is not a simple power-law. However, the TeV spectrum in model B is a simple power-law and extends into higher energy than that in model A. For possible and different redshifts (e.g., $z\lesssim1.0$), the nearly same intrinsic TeV spectrum in each model can be obtained by using slightly different parameters. In a specific model, this intrinsic TeV spectrum is constrained by the GeV spectrum, which is independent of the redshift.

Secondly, using the intrinsic TeV spectra we fit the observed TeV spectrum after taking the uncertainties on the redshift and EBL density into account. For the EBL uncertainties, we introduce a scaling factor $\tau_{\rm f}$, and define $\tau(z,\epsilon_1)=\tau_{\rm f}\cdot\tau(z,\epsilon_1)^{\rm model}$, where $\tau(z,\epsilon_1)^{\rm model}$ is the optical depth predicted by the EBL models.
According to the results given by \citet{Ackermann2012}, we restrict $\tau_{\rm f}$ between 0.6 and 1.0 for the EBL model of \citet{Finke10}. In model B, in addition to the EBL absorption, an internal absorption due to the low energy photon field around the jet is needed. Here, the low energy photon field is described as a diluted black body spectrum (see the description of model B in Section 2). In our fitting, we set $U_0$ to be $2.0\times10^{-5}\ \rm erg\ cm^{-3}$, and take $T$ and $R$ as free parameters. $T$ and $L$ calculated by using $U_0$ and $R$ define the feature of the soft photon field.
The fit results are listed in Table~\ref{TeV}, and two-dimensional $z-\tau_{\rm f}$ 68\% and 95\% confidence level contours are shown in Fig.~\ref{limit}. Then, we use the fit results in Table~\ref{TeV} to correct the observed TeV spectrum, which are plotted in Fig.~\ref{SED}. To fit TeV spectrum well, model A requires smaller EBL density and redshift compared with model B (see Table~\ref{TeV}). With the EBL density very close to the lower limit given by \citet{Ackermann2012} for \citet{Finke10} EBL model, the allowed 68\% redshift upper limits for models A and B are 0.75 and 1.03, respectively (see Table~\ref{TeV}).

In Fig.~\ref{neutrino}, we show the neutrino spectra calculated in the two models. Compared with model B, because of the much higher $p\gamma$ interaction rate in model A, the calculated neutrino flux in model A is significantly larger than that in model B. It is noted that the neutrino flux in model A is very close to the three years' sensitivity of the Askaryan Radio Array. Moreover, we also note that the shapes of the neutrino spectra predicted in the two models are different, which could be attributed to different values of the maximum proton energy in models A and B.

We now pay attention to the rest of differences between the two models. We note that the energy in the relativistic protons nearly equals to the energy in the magnetic field in model A (Table~\ref{modelp}, $U_p/U_B=1.4$) and the power in the relativistic protons is $L_{\rm p}=2.4\times10^{46} \rm\ erg\ s^{-1}$ (with $z=0.6$). While in model B the ratio of the energy in the relativistic protons to that in the magnetic field is far smaller than unit, which is caused by the dramatic decrease of the injected proton density (and then results in the decrease of $p\gamma$ interaction rate). In model B, the required power in the relativistic protons, $L_{\rm p}=1.3\times10^{45} \rm\ erg\ s^{-1}$ (with $z=0.6$), is one order of magnitude lower than that in model A. In model B, a higher maximum energy is used, which is still consistent with the Hillas condition.
The maximum proton energies in the blazar jet are discussed in \citet{murase12,murase14} and \citet{Dermer14}, and the value we used in model B is very close to the upper limit given in \citet{murase14} and \citet{Dermer14}.

Finally, let us discuss the reasonableness of the soft photon field required in model B although
we gave a reasonable assumption on it.
According to the values of $T$ and $R$ ($T=1.09^{+0.55}_{-0.19}\ $eV and $R=2.09^{+0.17}_{-0.69}\ $pc) derived from our fit and $U_0=2\times10^{-5}\rm \ erg\ cm^{-3}$ we assumed, it seems that this photon field comes from a hot dusty torus. But its luminosity within $2\ $pc from central black hole, $L\sim10^{43}\rm\ erg\ s^{-1}$, is extremely high for HBL. \citet{Stocke} reported Ly$\alpha$ luminosity of\ $\leq10^{41}\rm\ erg\ s^{-1}$ for three typical high-synchrotron-peak BL Lacs (HBLs) (Mrk 421, Mrk 501, and PKS 2005-489). Based on the current knowledge on blazars \citep[e.g.,][]{ghisellini09}, it is unlikely for PKS 1424+240 to have a soft photon field around the jet with $L\sim10^{43}\rm\ erg\ s^{-1}$.

\section{Conclusion and discussion}
\label{cd}

PKS 1424+240 may be the most distant TeV blazar at present. Understanding its intrinsic TeV spectrum generated in the jet is
very important, which could connect to some new physical or astrophysical phenomena \citep[e.g.,][]{Rubtsov,Tavecchio14}.
In this work, including the uncertainties of redshift and EBL density we have studied the origin of the TeV emission from PKS 1424+240 with two scenarios in the frame of the lepto-hadronic jet model. In the first scenario, TeV emission is attributed to the synchrotron radiation of UHE-photon-induced pair cascade and GeV emission is attributed to the proton synchrotron emission (model A). In the second scenario, GeV-TeV emission is attributed to the proton-synchrotron emission and an internal absorption is included (model B). Our results show that both models A and B are able to explain the TeV emission of PKS 1424+240 well with the redshift $z\lesssim0.7$ and a low EBL density ($\tau_{\rm f}\sim0.6$ for \citet{Finke10} EBL model). The allowed 68\% upper limits of the redshift for models A and B are 0.75 and 1.03, respectively. However, the required soft photon field in model B is unlikely for HBLs. Therefore, we could exclude the case of model B here for PKS 1424+240 although the parameters on the proton energy distribution and emission region are reasonable. We therefore conclude that model A can explain the TeV spectrum of PKS 1424+240 with a low EBL density when $0.6<z<0.75$.

We would like to stress that the proton-synchrotron emission producing GeV-TeV spectra is still reasonable for PKS 1424+240. \citet{Aleks2014} showed that {\it Fermi}-LAT spectrum connects smoothly with the MAGIC spectra in 2010 and 2011 (de-absorbed spectra assuming $z=0.6$ and with \cite{Franceschini} EBL model), which could be modeled well by the proton-synchrotron emission.
Based on the results presented in \citep{Aleks2014} (their Fig.~8), however, we estimate that {\it Fermi}-LAT spectrum and the MAGIC spectra in 2010 and 2011 could not be modeled well by the two-zone SSC model suggested in \citet{Aleks2014} and the proton-synchrotron emission if the redshift $z>0.6$.

For the VERITAS spectrum in 2009, our results have shown that it can be explained reasonably by model A with a low EBL density
if $z<0.75$ (68\% upper limit). Combining with the above discussions, this means that if $z>0.75$ the jet models could not account for the TeV emission from PKS 1424+240 because of the very serious EBL absorption. In this case ($z>0.75$), the non-jet models could explain the TeV spectra of PKS 1424+240, where the EBL absorption effect is reduced significantly so that no peculiarly hard TeV spectrum occurs. These non-jet models include the secondary emission produced during the propagation of UHECRs \citep[e.g.,][]{essey14} and the photon-axion-like particles oscillation models \citep[e.g.,][]{Meyer}.
Our study of the TeV spectrum formed in the jet would put some constraints on the non-jet models, which will be done in our coming work.

\section*{Acknowledgments}
We are grateful to Professor Shuangnan Zhang for carefully reading the manuscript and suggestions. We thank Amy Furniss for sending us the multiwavelength data sets. This work is partially supported by the National Natural Science Foundation of China (NSFC 11433004) and the support of Science Foundation for graduate students of Yunnan Province Education Department under grant no. 2013J071.

\bibliography{refernces}

\end{document}